\documentclass[a4paper,11pt]{article}
\usepackage{pos}
\usepackage{subfigure, upgreek}
\usepackage{amsmath}
\usepackage{nicefrac}
\usepackage[utf8]{inputenc}

\everymath{\displaystyle}
\DeclareMathAlphabet{\mathcal}{OMS}{cmsy}{m}{n}
\usepackage[absolute,overlay]{textpos}

\usepackage{lipsum}

\let\OLDthebibliography\thebibliography
\renewcommand\thebibliography[1]{
  \OLDthebibliography{#1}
  \setlength{\parskip}{0pt}
  \setlength{\itemsep}{4.1pt plus 0.3ex}
}

\title{$\pi\pi$ scattering at Large $N_\text{c}$}

\author*[a]{Jorge Baeza-Ballesteros}
\affiliation[a]{IFIC, CSIC-Universitat de València,\\ 46980 Paterna, Spain.}
\affiliation[b]{Center for Theoretical Physics, Massachusetts Institute of Technology, \\ Cambridge, MA 02139, USA.}
\emailAdd{jorge.baeza@uv.es}

\author[a]{Pilar Hernández}

\emailAdd{m.pilar.hernandez@uv.es}

\author[a,b]{Fernando Romero-López}

\emailAdd{fromerol@mit.edu}

\abstract{We study the Large $N_\text{c}$ scaling of pion-pion scattering amplitudes for $N_\text{f}=4$ degenerate quark flavors. We focus on the standard isospin-2 channel and the adjoint-antisymmetric (AA) representation. The latter only exists for $N_\text{f}\geq 4$ and a representative state is $\frac{1}{\sqrt{2}}(|D_s^+\pi^+\rangle-|D^+ K^+\rangle)$.  We compare the results obtained for two regularizations (Wilson and twisted-mass fermions) and three values of the lattice spacing, and observe significant discretization effects in the AA channel. Finally, we match our results to NLO SU(4) and NNLO U(4) Chiral Perturbation Theory and constrain the $N_\text{c}$ scaling of the relevant low-energy couplings.
}

\FullConference{%
 The 38th International Symposium on Lattice Field Theory, LATTICE2021
  26th-30th July, 2021
  Zoom/Gather@Massachusetts Institute of Technology \\
  Report number: IFIC/21-42, MIT-CTP/5345
}

\begin{document}
\maketitle


\section{Introduction}
\label{sec:intro}

The 't Hooft limit of QCD \cite{tHooft:1973alw}, that is the limit of large number of colors, $N_\text{c}$, is a simplification of the theory of strong interactions. It captures most of the non-perturbative features of QCD, such as asymptotic freedom, spontaneous chiral symmetry breaking, confinement, or the existence of a low energy spectrum of pseudo-Goldstone bosons. Moreover, it has proven to have predictive power in the non-perturbative regime, and is often used by phenomenological approaches to QCD. 

Several studies have addressed the Large $N_\text{c}$ limit via lattice simulations \cite{ReviewPilar}. Particularly interesting are questions where Large $N_\text{c}$ predictions seem to fail, such as non-leptonic kaon decays. Intrinsic QCD effects in these processes at Large $N_\text{c}$ have been recently studied by our group \cite{Donini:2020qfu}, finding that subleading $N_\text{c}$ corrections can naturally account for the discrepancy. However, final state interactions might also be relevant and remain to be studied. Another interesting open question is whether exotic states such as tetraquarks survive in the Large $N_\text{c}$ limit. Both these questions can be explored studying scattering processes on the lattice. 

\begin{figure}[h!]
   \centering
   \subfigure%
             {\includegraphics[width=0.27\textwidth,clip]{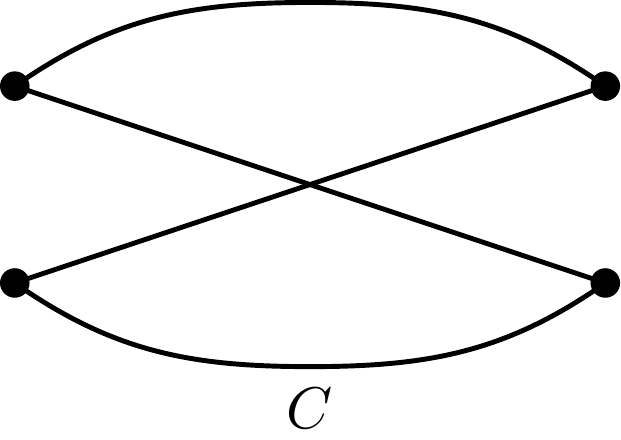}}\hspace{1.5 cm}
   \subfigure%
             {\includegraphics[width=0.27\textwidth,clip]{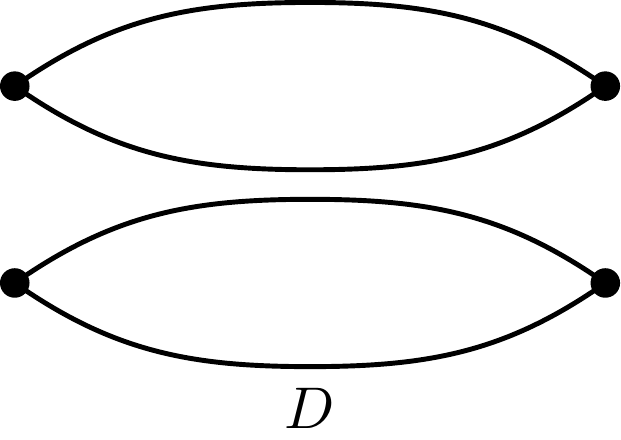}}
   \caption{Representation of the quark contractions required to compute the 2-pion correlators in the lattice for the $I=2$ and $AA$ channels.}
   \label{fig:contractions}
\end{figure}

In this talk we report the current state of our study of $\pi\pi$ scattering at Large $N_\text{c}$. We work in a theory with $N_\text{f}=4$ degenerate quark flavors, for which seven different irreducible representations (irreps) of SU(4)$_\text{f}$ exist \cite{Bijnens}. We have focused on two of them:
\begin{itemize}
\item The 84-dimensional irrep, which is the analogous to the isospin-2 ($I=2$) channel of SU(2). A representative state for this channel is the well-known $|\pi^+\pi^+\rangle$.
\item The 20-dimensional irrep, which only exists for $N_\text{f}\geq 4$. It is also known as the adjoint-antisymmetric ($AA$) channel. A representative state is $\frac{1}{\sqrt{2}}(|D_s^+\pi^+\rangle-|D^+ K^+\rangle)$.
\end{itemize}
for these two channels, two-pion correlation functions are computed on the lattice as linear combinations of the disconnected and connected quark contractions (see Fig. \hspace{-0.05cm}\ref{fig:contractions}):
\begin{equation}
C_{I=2}=2D-2C,\hspace{2cm} C_{AA}=2D+2C.
\end{equation}
From the lattice results, we extract scattering properties and match them to Chiral Perturbation Theory to constrain the Large $N_\text{c}$ scaling of the relevant Low Energy Couplings.

\section{$\pi\pi$ scattering in ChPT}
\label{sec:ChPT}

Chiral Perturbation Theory (ChPT) is an effective theory that describes the low-energy behavior of QCD in terms of the lightest non-singlet multiplet of mesons and a finite number of Low Energy Couplings (LECs). These mesons are the pseudo-Goldstone bosons resulting from the pattern of spontaneous chiral symmetry breaking of QCD, SU($N_\text{f})_\text{L}\times$SU($N_\text{f})_\text{R}\rightarrow$SU$(N_\text{f})_\text{V}$. ChPT has been widely studied \cite{Weinberg, Leutwyler} and in the case of degenerate quarks, $\pi\pi$ scattering amplitudes are known up to next-to-next-to-leading-order (NNLO) \cite{Bijnens}. From these results, the ($s$-wave) scattering lengths for $N_\text{f}=4$ and at NLO can be extracted:
\begin{equation}
M_{\pi}a_0^{I=2}=-\frac{M_\pi^2}{16\uppi F_{\pi}^2}\left[1-\frac{16 M_{\pi}^2}{F_{\pi}^2}L_{I=2}+\frac{M_{\pi}^2}{32\uppi^2 F_{\pi}^2}\left(\frac{13}{4}\ln{\frac{M_{\pi}^2}{\mu^2}}-\frac{3}{4}\right)\right],
\end{equation}
\begin{equation}
M_{\pi}a_0^{AA}=\frac{M_\pi^2}{16\uppi F_{\pi}^2}\left[1-\frac{16 M_{\pi}^2}{F_{\pi}^2}L_{AA}-\frac{M_{\pi}^2}{32\uppi^2 F_{\pi}^2}\left(\frac{21}{4}\ln{\frac{M_{\pi}^2}{\mu^2}}+\frac{5}{4}\right)\right],
\end{equation}
where $M_\pi$ and $F_\pi$ are the pion mass and decay constants, respectively, and $L_{I=2}$  and $L_{AA}$ are linear combinations of LECs. We can expand them as a power series in $N_\text{c}$,
\begin{equation}
L_{I=2}=N_\text{c}L^{(0)}+L^{(1)}_{I=2}+\mathcal{O}(N_\text{c}^{-1}), \hspace{1.5cm}
L_{AA}=N_\text{c}L^{(0)}+L^{(1)}_{AA}+\mathcal{O}(N_\text{c}^{-1}).
\end{equation} 
Note that the leading dependence of both quantities is expected to be the same.

In the 't Hooft limit, we must include singlet meson, the $\eta'$, in the effective theory. Its mass originates from the explicit U(1)$_\text{A}$ breaking by the anomaly. This is suppressed at Large $N_\text{c}$, meaning it becomes degenerate with the rest of mesons,
\begin{equation}
M_{\eta'}^2=M_\pi^2+\frac{2N_\text{f}\chi_\text{top}}{F_\pi^2}\xrightarrow{\text{Large }N_\text{c}}M_\pi^2+\mathcal{O}(N_\text{c}^{-1}).
\end{equation}
where $\chi_\text{top}$ is the topological susceptibility of pure Yang-Mills and $F_\pi^2\sim\mathcal{O}(N_\text{c})$.

An extension of ChPT to Large $N_\text{c}$ has already been studied and is sometimes referred to as Large $N_\text{c}$ or U($N_\text{f}$) ChPT \cite{LargeNChPT}. It includes the $\eta'$ in the pion matrix and $N_\text{c}$ in the counting scheme, $\mathcal{O}(m_q)\sim\mathcal{O}({M_\pi^2})\sim\mathcal{O}({p^2})\sim\mathcal{O}({N_\text{c}^{-1}})$, with $m_q$ the quark mass and $p$ the external momentum.  As a consequence, loop diagrams first enter the computations at NNLO.

Within this theory, we have computed the scattering amplitudes for both channels of interest to NNLO. For $N_\text{f}=4$ the scattering lengths are:
\begin{equation}
\begin{array}{rl}
M_{\pi}a_0^{I=2}=-\frac{M_\pi^2}{16\uppi F_{\pi}^2}&\left\{1-\frac{16 M_{\pi}^2}{F_{\pi}^2}L_{I=2}+K_{I=2}\left(\frac{M_\pi^2}{F_{\pi}^2}\right)^2\right. \\
&\left.
+\frac{M_{\pi}^2}{32\uppi^2 F_{\pi}^2}\left[\frac{15M_\pi^2-13M_{\eta'}^2}{4(M_\pi^2-M_{\eta'}^2)}\ln{\frac{M_{\pi}^2}{\mu^2}}+\frac{M_\pi^2-3M_{\eta'}^2}{4(M_\pi^2-M_{\eta'}^2)}\ln{\frac{M_{\eta'}^2}{\mu^2}}-\frac{1}{2}\right]\right\},
\end{array}
\end{equation}
\begin{equation}
\begin{array}{rl}
M_{\pi}a_0^{AA}=\frac{M_\pi^2}{16\uppi F_{\pi}^2}&\left\{1-\frac{16 M_{\pi}^2}{F_{\pi}^2}L_{AA}+K_{AA}\left(\frac{M_\pi^2}{F_{\pi}^2}\right)^2 
\right.\\
&\left. -\frac{M_{\pi}^2}{32\uppi^2 F_{\pi}^2}\left[\frac{15M_\pi^2-21M_{\eta'}^2}{4(M_\pi^2-M_{\eta'}^2)}\ln{\frac{M_{\pi}^2}{\mu^2}}+\frac{M_\pi^2+5M_{\eta'}^2}{4(M_\pi^2-M_{\eta'}^2)}\ln{\frac{M_{\eta'}^2}{\mu^2}}+\frac{3}{2}\right]\right\},
\end{array}
\end{equation}
where $K_{I=2}$ and $K_{AA}$ are combinations of SU($N_\text{f}$) LECs and new ones from the U($N_\text{f}$) theory. They scale as $\mathcal{O}(N_\text{c}^2)$ and only the leading dependence is considered as a fitting parameter when matching to lattice results.

\begin{figure}[h!]
   \centering
\includegraphics[width=0.57\textwidth,clip]{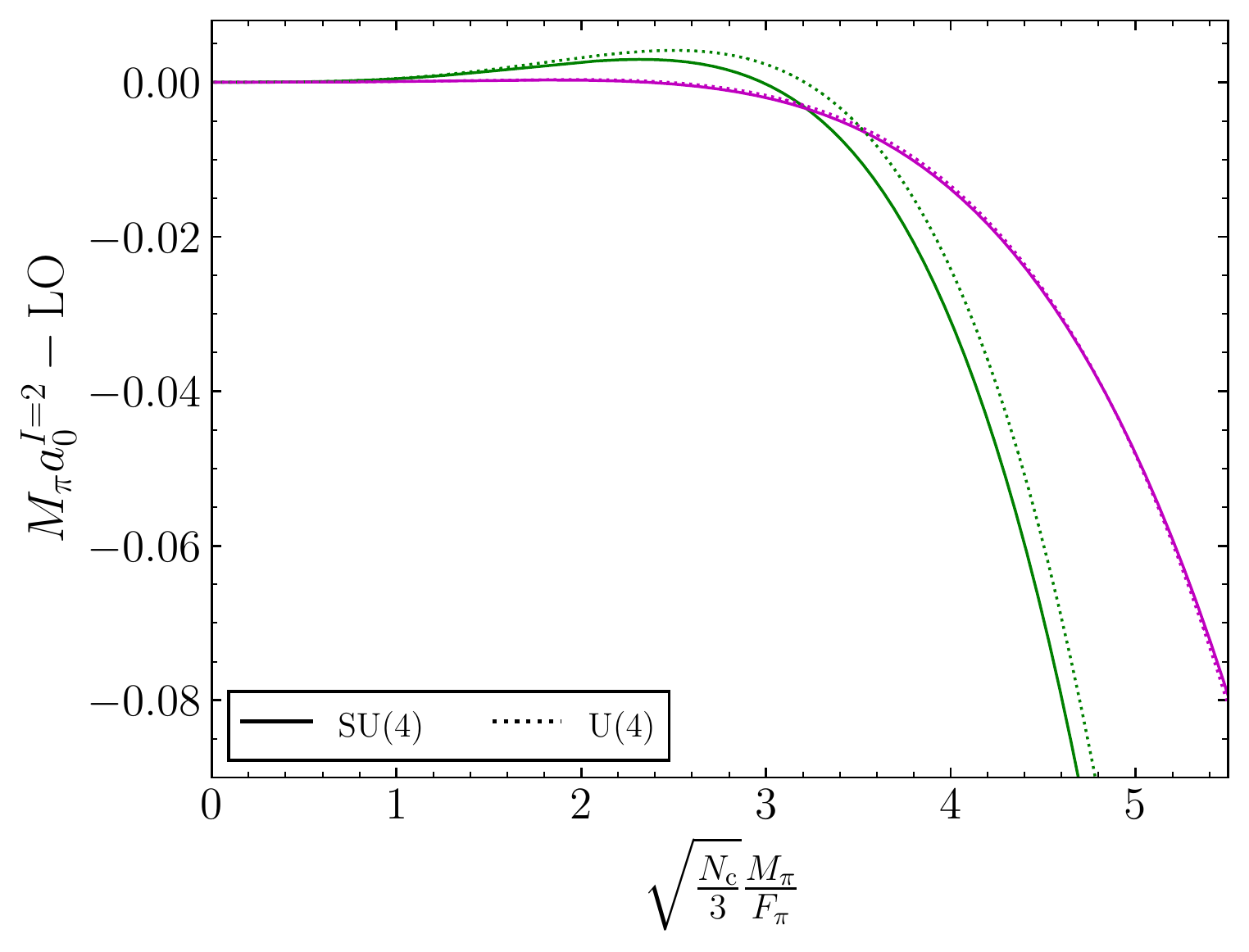}
   \caption{Comparison between the SU(4) and U(4) ChPT one-loop prediction for the scattering length in the $I=2$ channel for $N_\text{c}=3$ (green) and 6 (pink). We have set $K_{I=2}=0$ and matched $L^{(1)}_{I=2}$ between both theories using the $M_{\eta'}\gg M_\pi$ limit.}
   \label{fig:Comparison}
\end{figure}

One can then compare these results with the ones from SU(4) ChPT. In Fig. \ref{fig:Comparison} we represent the scattering length in the $I=2$ channel for both theories and two values of $N_\text{c}$. We have set $K_{I=2}=0$ for the comparison and matched the values of the LECs between both theories using the $M_{\eta'}\gg M_\pi$ limit. The differences observed between both theories vary with $N_\text{c}$ as a result of two effects. First, the contribution from chiral logarithms including the $\eta'$ increases as $M_{\eta'}$ approaches $M_\pi$, but at the same time, the importance of the NNLO corrections gets suppressed with $N_\text{c}$.

\section{Finite volume formalism}
\label{sec:finotevol}

Lattice simulations allow to obtain the energy spectrum of two pions in a finite volume. One must then relate it to infinite volume scattering quantities, such as the scattering amplitude. This is done using Lüscher's formalism \cite{Luscher}. For the lowest energy state, the $s$-wave phase shift, $\delta_0$, can be related to the ground state energy of two pions, $E_{\pi\pi}$, on a cubic box of side $L$, as
\begin{equation}
k\cot\delta_0=\frac{1}{\uppi L}\mathcal{Z}\left(\frac{Lk}{2\uppi}\right)
\end{equation}
where $k$ is the center-of-mass (CM) momentum, defined from $E_{\pi\pi}=\sqrt{k^2+M_\pi^2}$, and $\mathcal{Z}$ is the generalized Lüscher's zeta function.

Near threshold, the energy shift of the ground state, $\Delta E_{\pi\pi}=E_{\pi\pi}-2M_\pi$, can be expanded in terms of the scattering length, $a_0$, and the effective range, $r_0$, of the system. 
\begin{equation}\label{eq:threshold}
\displaystyle{\Delta E_{\pi\pi}=-\frac{4\uppi a_0}{M_\pi L^3}\left[1+c_1\left(\frac{a_0}{ L}\right)+c_2\left(\frac{a_0}{L}\right)^2+ c_3\left(\frac{a_0}{L}\right)^3 +\frac{2\uppi r_0a_0}{L^3}+\frac{\uppi a_0}{M_\pi^2 L^3} \right]},
\end{equation}
where $c_1$, $c_2$ and $c_3$ are known constants. This result was first developed up to $\mathcal{O}(L^{-5})$ in Ref. \cite{Luscher}, and the order $\mathcal{O}(L^{-6})$ was later worked out in Ref. \cite{Steve}.

\section{Energy spectrum from lattice simulations}
\label{sec:energyspectrum}

Our lattice ensembles have been generated using HiRep \cite{DelDebbio}. We have 17 ensembles with $N_\text{f}=4$, $N_\text{c}=3-6$ and lattice spacing $a=0.075$ fm. We have also produced 4 finer ensembles for $N_\text{c}=3$, two with $a=0.065$ fm, and another two with $a=0.059$ fm, which are used to study discretization effects. We use the Iwasaki gauge action with $N_\text{f}=4$ clover-improved Wilson fermions on the sea, and two different regularizations for the valence sector:
\begin{itemize}
\itemsep0em 
\item A unitary setup with improved Wilson fermions.
\item A mixed-action setup at maximal twist.
\end{itemize}
Both setups are expected to show $\mathcal{O}(a^2)$ improvement and represent an extra handle  to analyze possible discretization effects, since they must coincide in the continuum. Moreover, the mixed-action setup allows us to compute $F_\pi$ from the one-pion correlator. A summary of the ensembles can be found in Refs. \cite{Donini:2020qfu, Pilar, Yo}.

\begin{figure}[h!]
   \centering
   \subfigure[$I=2$-channel energy shift for a $N_\text{c}=4$ ensemble]%
             {\includegraphics[width=0.478\textwidth,clip]{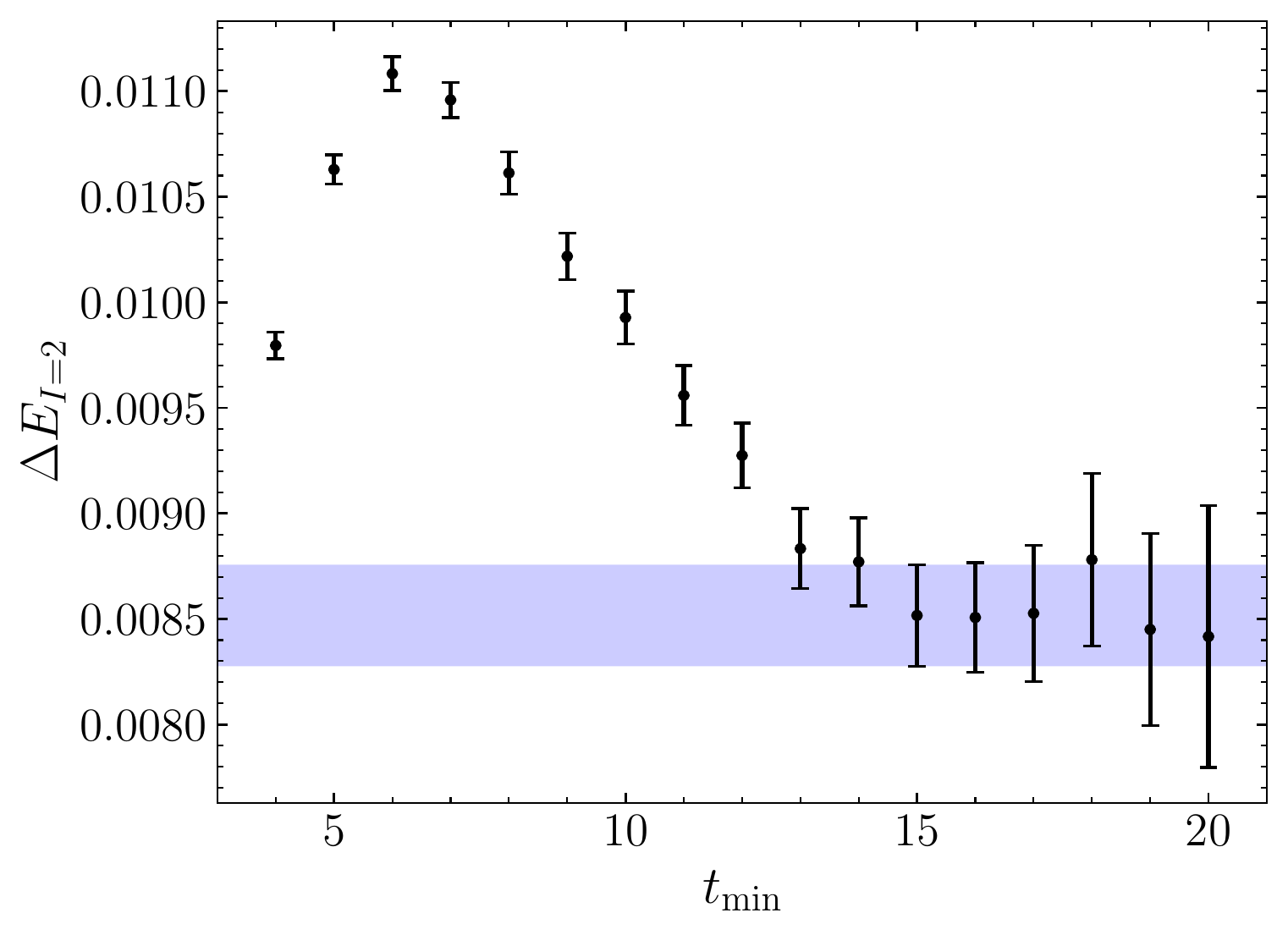}}\hfill
   \subfigure[$AA$-channel energy shift for a $N_\text{c}=6$ ensemble]%
             {\includegraphics[width=0.49\textwidth,clip]{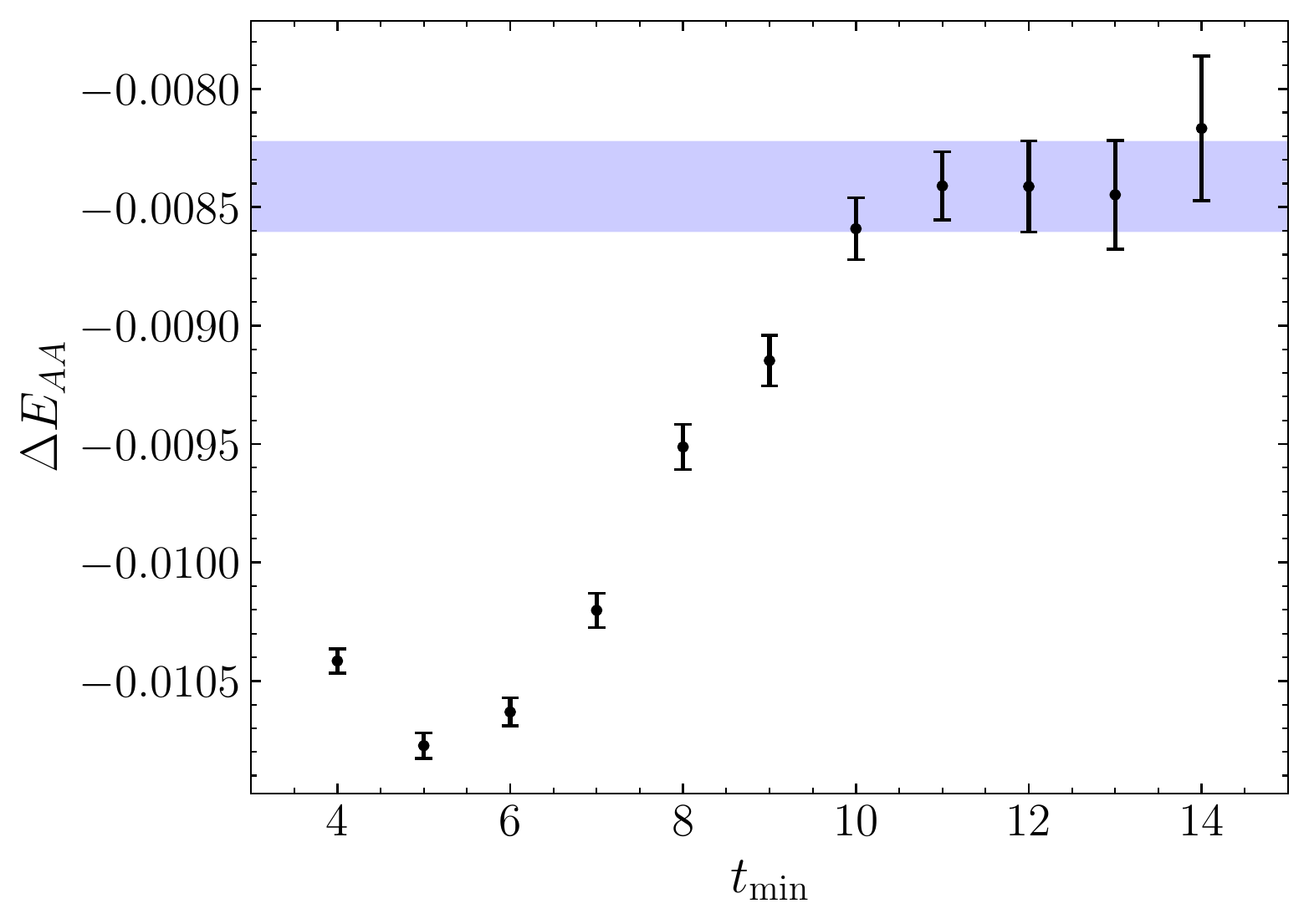}}
   \caption{Energy shifts obtained for different fit ranges. The final result is extracted from the plateau.}
   \label{fig:plateaus}
\end{figure}

We have computed the one- and two-pion correlators, $C_\pi$ and $C_{\pi\pi}$, respectively, in both channels. We then extract the ground state energy shift fitting to the following ratio function \cite{Feng}:
\begin{equation}
R(t)=\frac{C_{\pi\pi}(t+a)-C_{\pi\pi}(t-a)}{C_\pi^2(t+a)-C_\pi^2(t-a)} \xrightarrow[]{T/2>t\gg1} A_{\pi\pi}\left[\cosh(\Delta E_{\pi\pi} t')+\sinh(\Delta E_{\pi\pi} t')\coth(2M_\pi t')\right],
\end{equation}
where $A_{\pi\pi}$ is a dimensionless amplitude, $t'=t-T/2$ and $T$ is the time extent of the lattice. Two examples of the results of the fits are shown in Fig. \ref{fig:plateaus} for different fitting ranges. The energy shifts are then extracted from where they form a plateau.

We now study discretization effects for $N_\text{c}=3$. In Fig. \ref{fig:discreteenery} we compare between both regularizations of the valence sector.  We observe that discretization effects are small for the $I=2$ channel (left), while they are as big as $\sim 50\%$ for the coarser ensembles in the $AA$ channel (right), and are reduced for decreasing $a$. To better understand these effects, we perform a continuum extrapolation for $N_\text{c}=3$ in the $AA$ channel. We use $k\cot\delta_0$ as our physical observable for the analysis and proceed in three steps:

\begin{figure}[h!]
   \centering
   \subfigure[$I=2$ channel]%
             {\includegraphics[width=0.49\textwidth,clip]{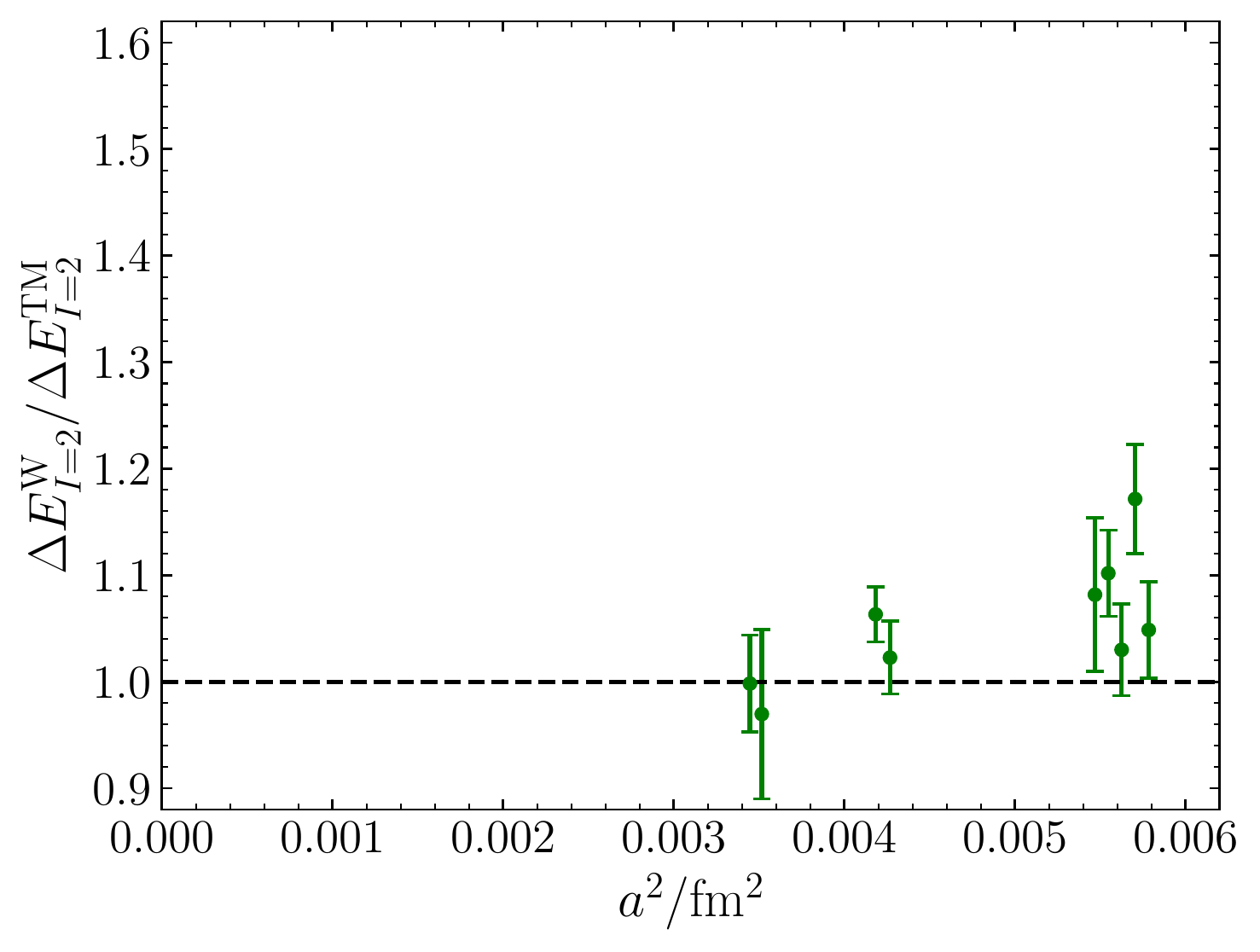}}\hfill
   \subfigure[$AA$ channel]%
             {\includegraphics[width=0.49\textwidth,clip]{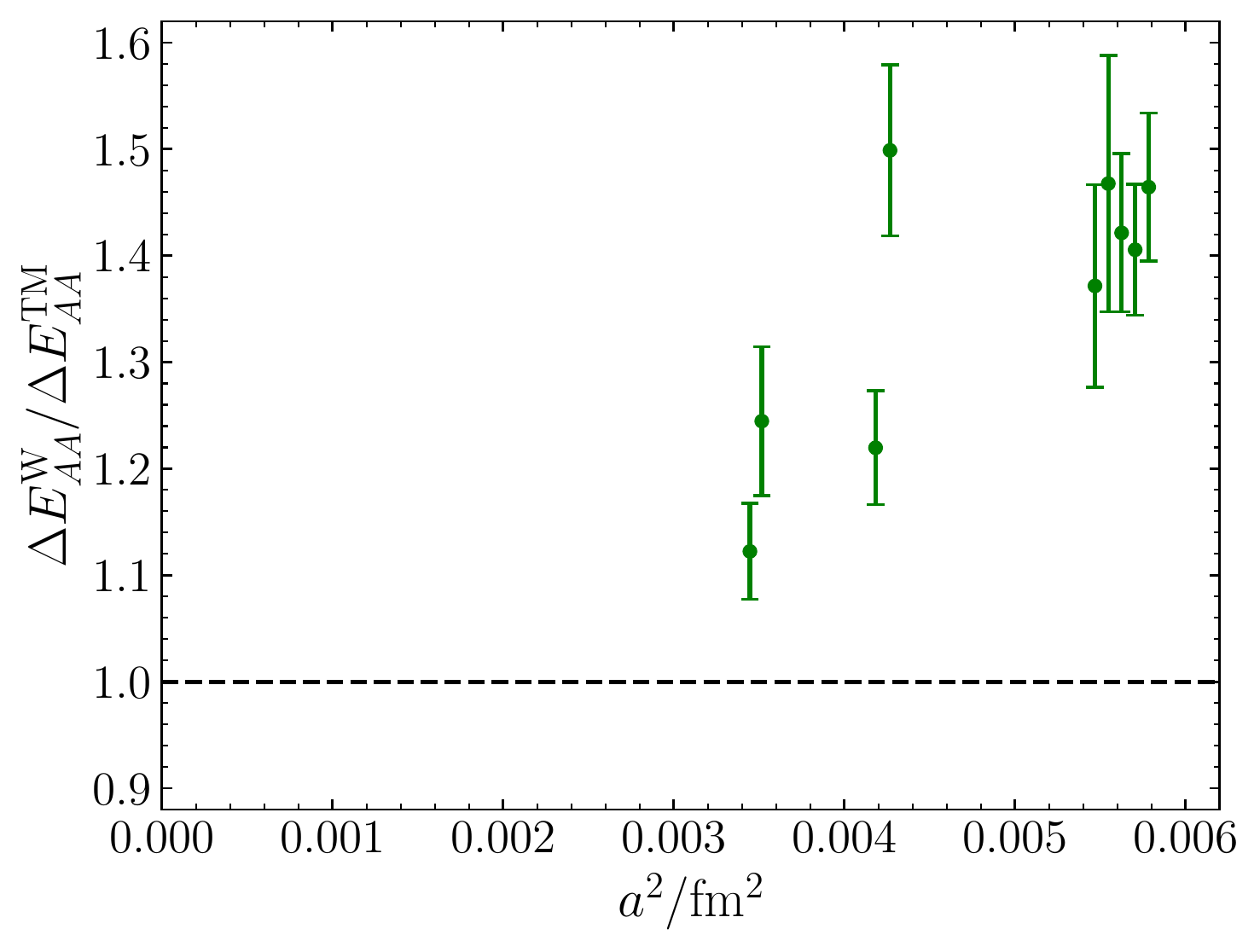}}
   \caption{Dependence on the lattice spacing of the ratio between the energy shifts obtained for the unitary Wilson and the mixed-action setups. }
   \label{fig:discreteenery}
\end{figure}

\begin{figure}[h!]
   \centering
\includegraphics[width=0.6\textwidth,clip]{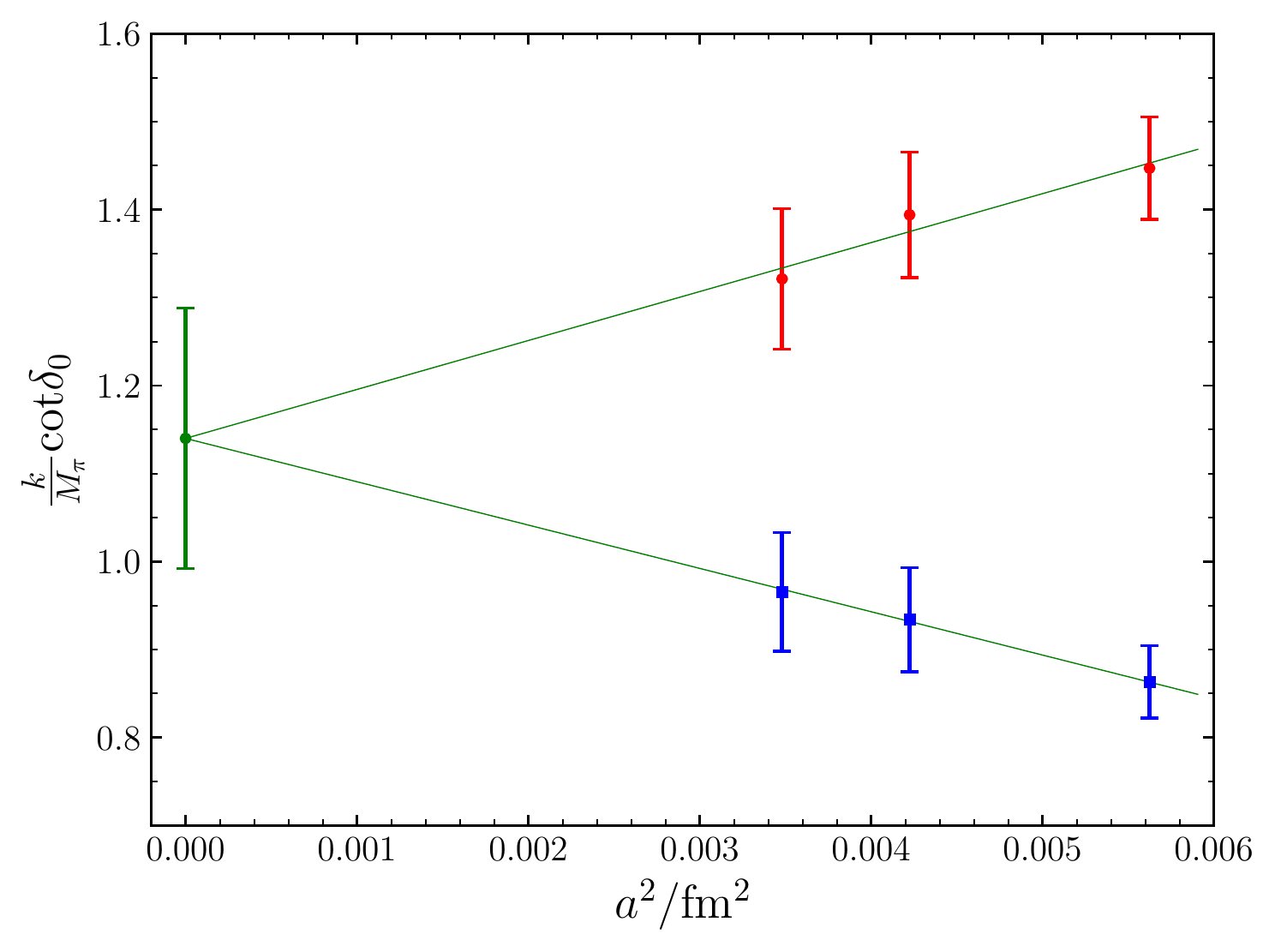}
   \caption{Continuum extrapolation of the $s$-wave phase shift for $N_\text{c}=3$. Results from the unitary setup (blue squares) and the mixed-action setup (red dots) are fitted simultaneously with a constrained common continuum limit.}
   \label{fig:extrapol}
\end{figure}

\begin{enumerate}
\item First, we extrapolate all data points to a fixed value of the momentum, $k/M_\pi=0.08$. We use the effective range expansion (ERE) and a prior for the effective range, $M_\pi^2a_0 r_0\in[-5,-1]$ motivated by the LO  prediction of ChPT, \vspace{-0.1 cm}
\begin{equation}\label{eq:LOChPT}
\left.M_\pi^2a_0 r_0\right|_\text{LO ChPT}=-3 \hspace{2cm} (I=2\text{ and }AA\text{ channel}).\vspace{-0.1 cm}
\end{equation}

\item For each lattice spacing, we interpolate to chiral parameter $\xi=M_\pi^2/(4\uppi F_\pi)^2=0.14$.
\item Finally we do a constrained continuum extrapolation, as shown in Fig. \ref{fig:extrapol}.
\end{enumerate}

We observe that our results are consistent with a universal continuum limit and that $\mathcal{O}(a^2)$ discretization effects are large for both regularizations. We decide to use the mixed-action results for the following analysis and parametrize the discretization effects in the scattering amplitudes as
\begin{equation}
\mathcal{M}_{AA}^\text{latt}=\mathcal{M}_{AA}^\text{cont}+  a^2 W\xi,
\end{equation}
which is inspired in a modification of ChPT that includes the effects of Wilson fermions and a twisted mass \cite{Buchoff, Stevetm}. Here, $W$ is a linear combination of new LECs appearing in this theory and obeys $W\sim \mathcal{O}(1)$ in $N_\text{c}$. It is treated as a fitting parameter when matching lattice results to ChPT.

\section{Fits to ChPT}
\label{sec:fits}

\begin{figure}[h!]
   \centering
   \subfigure[$I=2$ channel. Points to the right of the dashed line\newline are not considered for the fit.]%
             {\includegraphics[width=0.511\textwidth,clip]{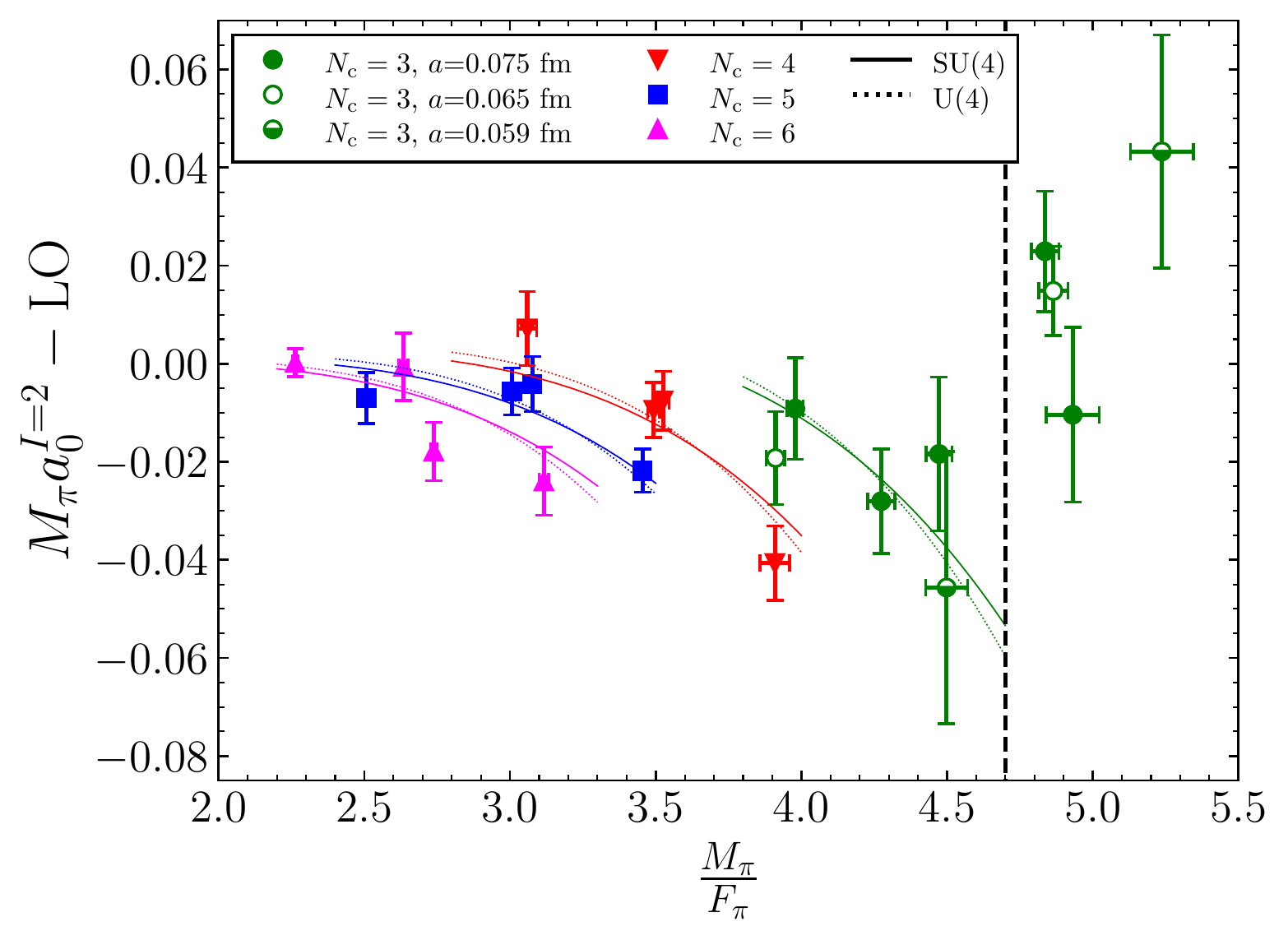}}\hfill
   \subfigure[$AA$ channel. Discretization effects are included as a fitting parameter.]%
             {\includegraphics[width=0.489\textwidth,clip]{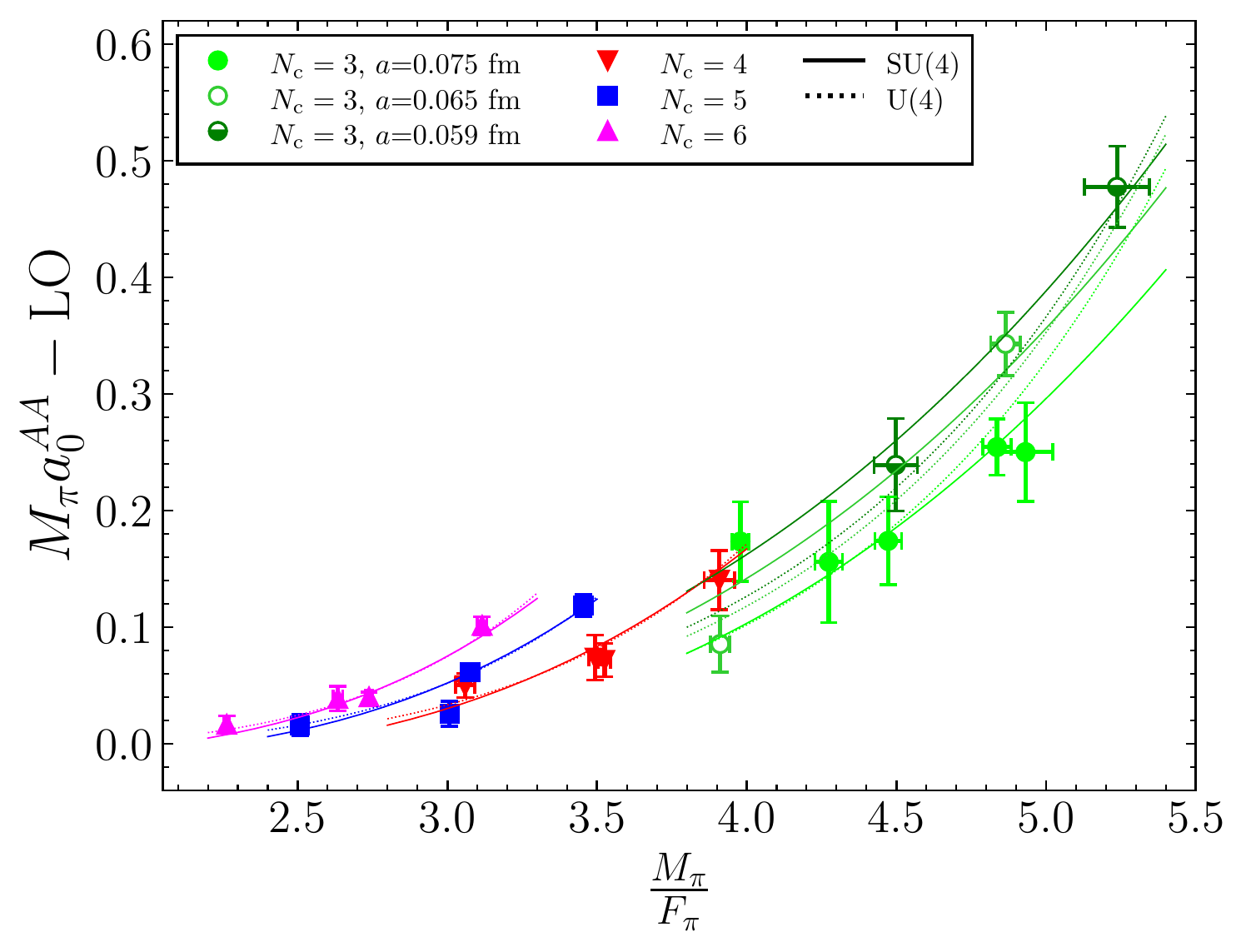}}
   \caption{Preliminary results for the simultaneous chiral and $N_\text{c}$ fits of the scattering length to SU(4) (solid line) and U(4) (dashed line) ChPT.}
   \label{fig:a0fit}
\end{figure}

We now compare our results to ChPT to constrain the $N_\text{c}$ scaling of the LECs. We use Eq. (\ref{eq:threshold}) to $\mathcal{O}(L^{-5})$ to extract $M_\pi a_0$ from the energy spectra and do a simultaneous chiral and $N_\text{c}$ fit to ChPT. The results are shown  in Fig.\hspace{-0.1cm} \ref{fig:a0fit} for both channels and both SU(4) and U(4) theories. In the $I=2$ channel, we observe that none of the theories explains the behavior of the most massive $N_\text{c}=3$ points, so we opt not to fit them. Our preliminary results for the LECs in the SU(4) theory are:
\begin{equation}
\arraycolsep=0pt\def\arraystretch{0}
\begin{array}{ll}
\displaystyle{L_{I=2}/N_\text{c}\times10^3 = -0.11(4)-1.43(16)/N_\text{c}}, \hspace{0.5 cm}& \chi^2/\text{dof}=1.00,\vspace{0.25 cm}\\

\displaystyle{L_{AA}/N_\text{c}\times10^3 = -1.08(13)+2.2(3)/N_\text{c}}, \hspace{0.5 cm}& \chi^2/\text{dof}=2.00,
\end{array}
\end{equation}
and in U(4) ChPT:
\begin{equation}
\arraycolsep=0pt\def\arraystretch{0}
\begin{array}{ll}
\displaystyle{L_{I=2}/N_\text{c}\times10^3 = -0.10(7)-1.29(16)/N_\text{c}}, \hspace{0.5 cm}& \chi^2/\text{dof}=0.94,\vspace{0.25 cm}\\

\displaystyle{L_{AA}/N_\text{c}\times10^3 = -0.6(4)+2.4(3)/N_\text{c}}, \hspace{0.5 cm}& \chi^2/\text{dof}=1.42.
\end{array}
\end{equation}
For both the SU(4) and U(4) ChPT fits, there is some discrepancy between the leading dependence of the LECs of the two channels, which were expected to be the same. 

\begin{figure}[h!]
   \centering\vspace{-0.2 cm}
   \subfigure[$I=2$ channel, $N_\text{c}=3$. ]%
             {\includegraphics[width=0.494\textwidth,clip]{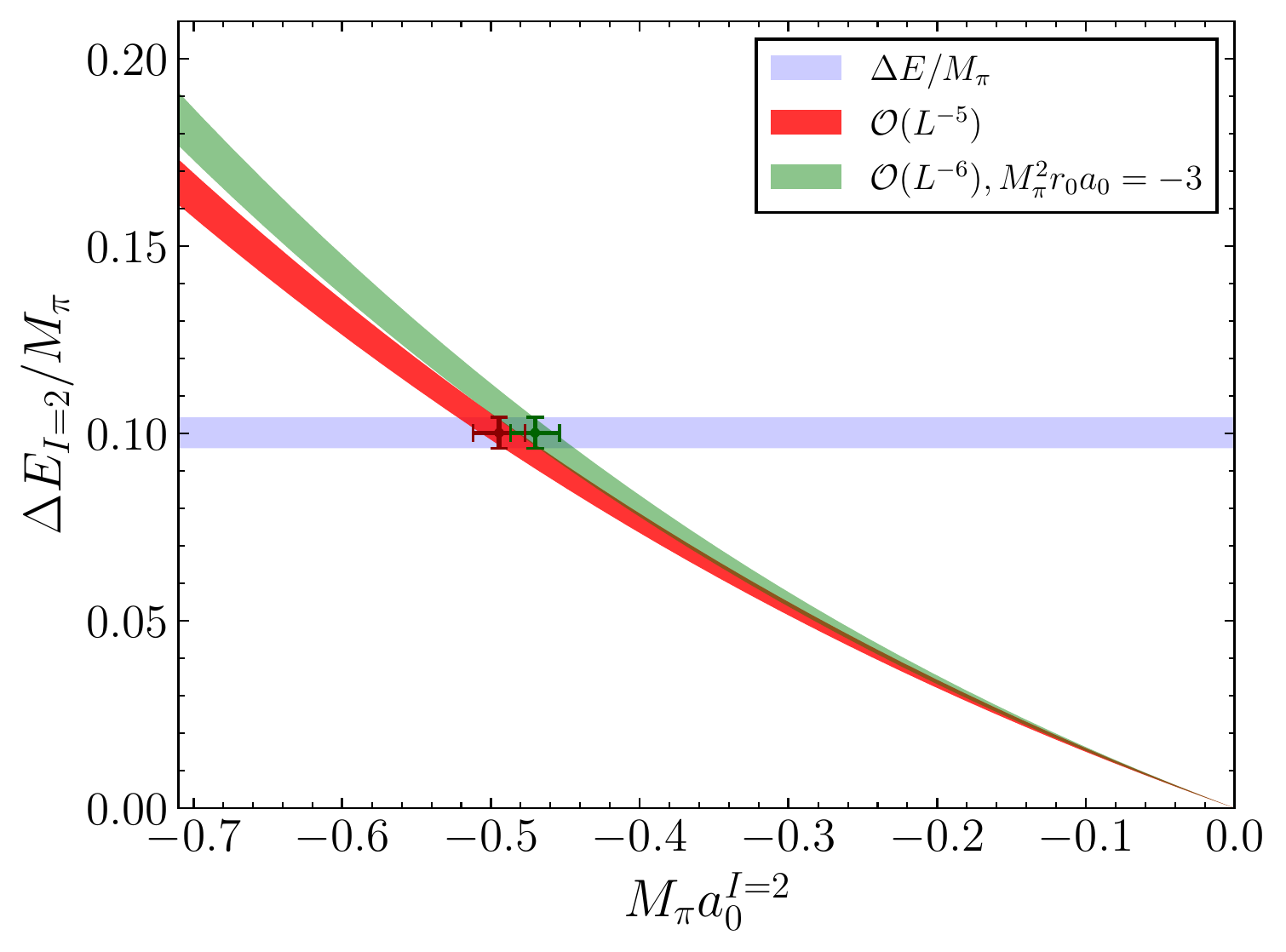}}\hfill
   \subfigure[$AA$ channel, $N_\text{c}=3$.]%
             {\includegraphics[width=0.5\textwidth,clip]{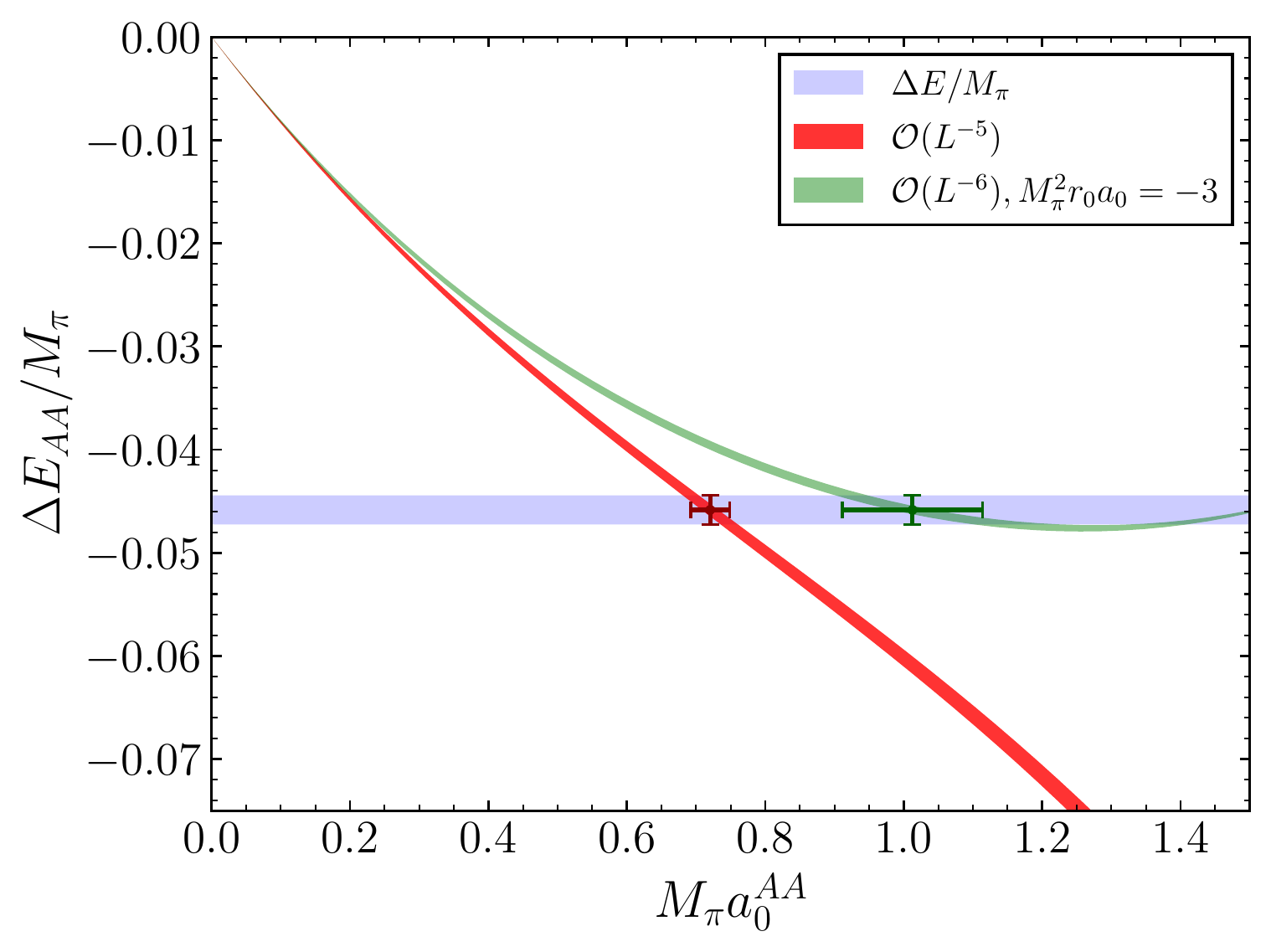}}
   \caption{Comparison between the $\mathcal{O}(L^{-5})$ (red) and $\mathcal{O}(L^{-6})$ (green) threshold expansion (Eq.\hspace{-0.1 cm} \ref{eq:threshold}). The horizontal blue region is the ground state energy shift from the lattice and points are the determinations of the scattering length.
}
   \label{fig:convergence}\vspace{-0.3 cm}
\end{figure}

We now study the impact of higher order terms in the threshold expansion. The analysis is represented for the two ensembles, one for each channel, in Fig. \ref{fig:convergence}. We compare the threshold expansion to $\mathcal{O}(L^{-5})$ (in red) and $\mathcal{O}(L^{-6})$ (in green) using Eq. \ref{eq:LOChPT} for $r_0$. The blue region depicts the energy shift from the lattice and the points are the determinations of $M_\pi a_0$. We observe that the $I=2$ channel shows good agreement between both truncation orders. However, this is not the case for the $AA$ channel for which convergence fails at large $\xi$ or small volume.

\begin{figure}[h!]
   \centering
   \subfigure[$I=2$ channel. Points to the right of the dashed line \newline are not considered for the fit.]%
             {\includegraphics[width=0.485\textwidth,clip]{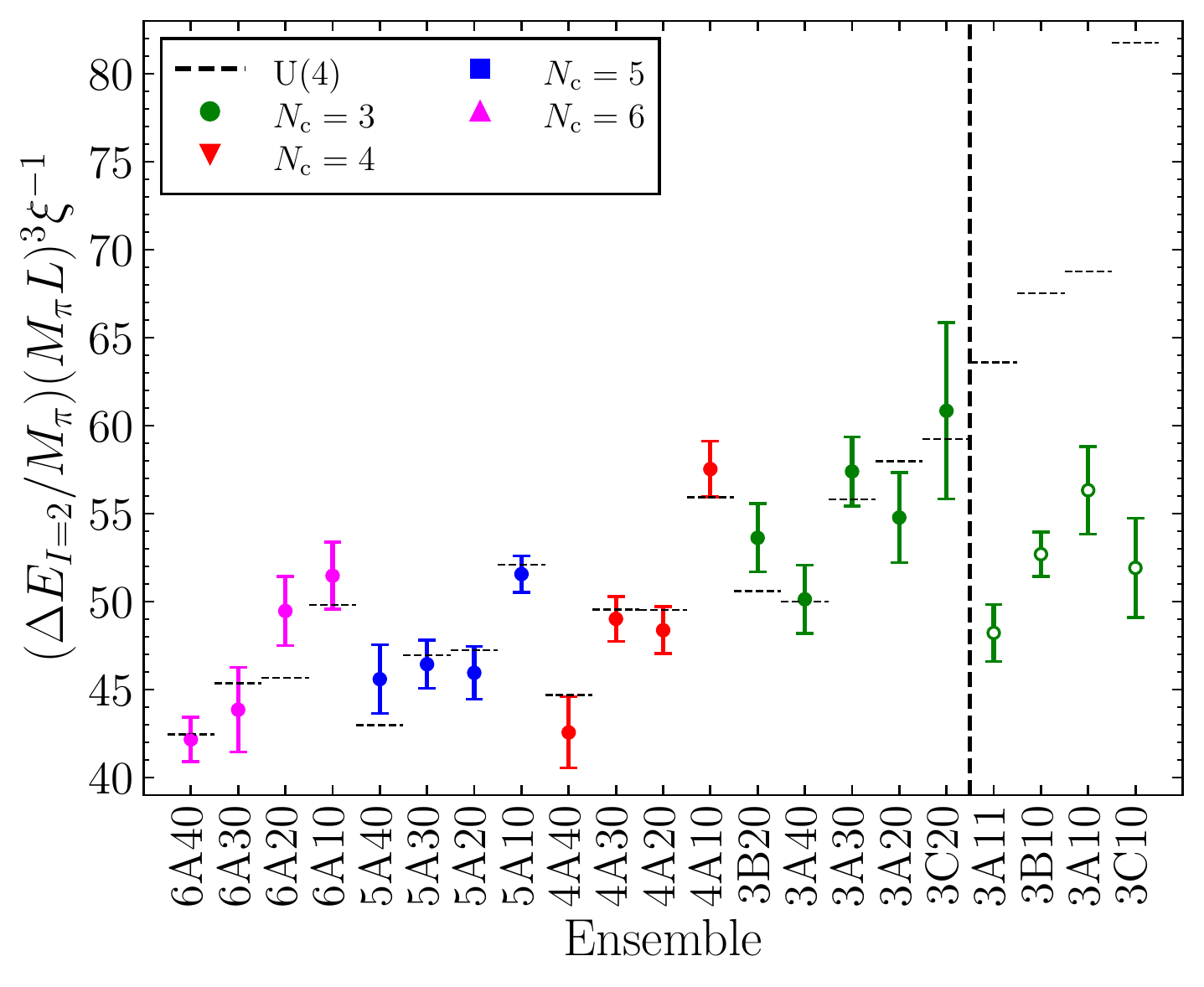}}\hfill
   \subfigure[$AA$ channel. Discretization effects are included as a fit parameter.]%
             {\includegraphics[width=0.513\textwidth,clip]{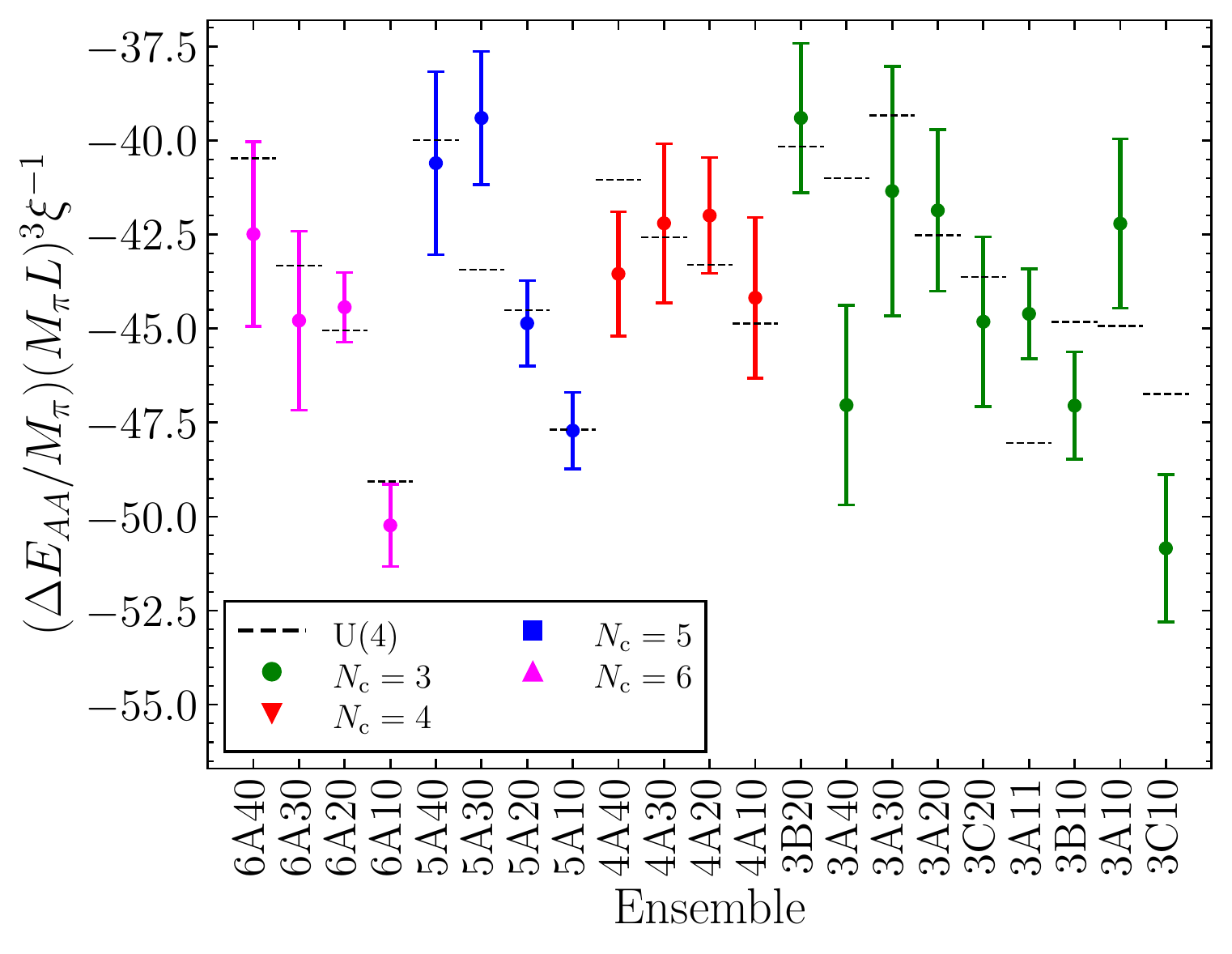}}
   \caption{Preliminary results for the simultaneous chiral and $N_\text{c}$ fits of the ground state energy shifts to U(4) ChPT.}
   \label{fig:energyfit}
\end{figure}

We thus decide to use the full Lüscher's formalism and do a simultaneous chiral and $N_\text{c}$ fit to the energy spectrum. The best fits to U(4) ChPT are shown in Fig. \ref{fig:energyfit} for both channels. From these we obtain the following preliminary results for the LECs,

 \begin{equation} 
\begin{array}{ll}
\displaystyle{L_{I=2}/N_\text{c}\times10^3 = -0.07(4)-1.4(2)/N_\text{c}}, \hspace{0.5 cm}& \chi^2/\text{dof}=0.85,\vspace{0.25 cm}\\

\displaystyle{L_{AA}/N_\text{c}\times10^3 = -0.9(2)+2.6(6)/N_\text{c}}, \hspace{0.5 cm}& \chi^2/\text{dof}=1.34.
\end{array}
\end{equation} 
These values are similar to the ones obtained before from the scattering length. Still, there is some tension between the leading $N_\text{c}$ contribution to the LECs.

\section{Summary and outlook}
\label{sec:summary}


In this talk, we have reported the current status of our study of $\pi\pi$ scattering at Large $N_\text{c}$. We have studied the $N_\text{c}$ scaling of the scattering amplitudes in the $I=2$ and $AA$ channels both numerically, and in U($N_\text{f}$) ChPT up to NNLO. From the comparison of both results, we have extracted the leading and subleading $N_\text{c}$ dependence of the relevant LECs. 

In future work, we intend to study other scattering channels and meson resonances. The $I=0$ is very appealing due to the presence of the $\sigma$ resonance and its contribution to final state interactions in the $K\rightarrow \pi\pi$ process. However, some preliminary work has shown it is much more computationally expensive. On the other hand, interactions in the $AA$ channel analyzed in this work are attractive, which may lead to the existence of exotic states at higher center-of-mass momentum. 

\section{Acknowledgments}
\label{sec:acknowledg}

This work has received support from the Generalitat Valenciana grant PROMETEO/2019/083, the European project H2020-MSCA-ITN-2019//860881-HIDDeN, and the national project FPA2017-85985-P. JBB is also supported by the Spanish grant FPU19/04326 of MCIU. FRL acknowledges funding from the European Union Horizon 2020 research and innovation program under the Marie Skłodowska-Curie grant agreement No. 713673 and "La Caixa" Foundation (ID 100010434, LCF/BQ/IN17/11620044). FRL has also received financial support from Generalitat Valenciana through the plan GenT program (CIDEGENT/2019/040). The work of FRL is supported in part by the U.S. Department of Energy, Office of Science, Office of Nuclear Physics, under grant Contract Numbers DE-SC0011090 and DE-SC0021006. We thank Mare Nostrum 4 (BSC), Finis Terrae II (CESGA), Cal\'endula (SCAYLE), Tirant 3 (UV) and Lluis Vives (Servei d'Informàtica UV) for the computing time provided. 



\begin{thebibliography}{99}


\bibitem{tHooft:1973alw}
  G.~'t Hooft,
  Nucl.\ Phys.\ B {\bf 72} (1974) 461.
  
\bibitem{ReviewPilar}
  	P. Hernández and F. Romero-López,
  	Eur. Phys. J. A {\bf 57} (2021)2, 52
  	[hep-lat/2012.03331].

\bibitem{Donini:2020qfu}
    A. Donini, P. Hernández, C. Pena and F. Romero-López,
    Eur. Phys. J. C {\bf 88}, no. 7, 638 (2020)
    [hep-lat/2003.10293].
    
\bibitem{Bijnens}
    J. Bijnens and J. Lu,
    JHEP {\bf 2011}, 28 (2011)
    [hep-ph/1102.0172].
    
\bibitem{Weinberg}
	S. Weinberg,
	Physica A {\bf 96} (1979) 1-2, 327-340.
	
\bibitem{Leutwyler}
	J. Gasser and H. Leutwyler,
	Nucl. Phys. B {\bf 250} (1985) 465-516.
	
\bibitem{LargeNChPT}
	R. Kaiser and H. Leutwyler,
	Eur. Phys. J. C {\bf 17} (2000) 623-649
	[hep-ph/0007101].
	
\bibitem{Luscher}
	M. Lüscher,
	Commun. Math. Phys. {\bf 105} (1986) 153-188.
	
\bibitem{Steve}
	M. T. Hansen and S. R. Sharpe,
	Phys. Rev. D {\bf 93} (2016) 014506
	[hep-lat/1509.07929].

\bibitem{DelDebbio}
	L. Del Debbio {\it et al.,}
	Phys. Rev. D {\bf 80} (2009) 074507 
	[hep-lat/0907.3896].
	


\bibitem{Pilar}
	P. Hernández, C. Pena and F. Romero-López,
	Eur. Phys. J. C {\bf79} (2019) 10, 865
	[hep-lat/1907.11511].
	
\bibitem{Yo}
	J. Baeza-Ballesteros, P. Hernández and F. Romero-López,
	in preparation (2021).	
	
\bibitem{Feng}
	X. Feng, K. Jansen and D. B.  Renner,
	Phys. Lett. B {\bf 684} (2010) 268-274 
	[hep-lat/0909.3255].
	
\bibitem{Buchoff}
	M.  I. Buchoff, J.-W. Chen and A. Walker-Loud,
	Phys. Rev. D {\bf 79} (2009) 074503 
	[hep-lat/0810.2464].
	
\bibitem{Stevetm}
	S. R. Sharpe and J. M. S. Wu,
	Phys. Rev. D {\bf 71} (2005) 074501 
	[hep-lat/0411021].

\end{thebibliography}
\end{document}